\documentclass[twocolumn]{aastex631}
\usepackage{natbib}

\usepackage{siunitx}
\usepackage{float}
\usepackage{physics}
\usepackage{amsmath}
\usepackage[caption = false]{subfig}
\usepackage{xcolor}
\usepackage{tablefootnote}

\shorttitle{TRIPP}
\shortauthors{}

\graphicspath{{./}{figures/}}

\begin{document}

\title{TRIPP: A General Purpose Data Pipeline for Astronomical Image Processing}

\email{agthomas@ucsb.edu}
\author[0000-0001-9528-8147]{Alex Thomas}
\affiliation{University of California Santa Barbara \\
Santa Barbara, CA 93106, USA}
\author{Natalie LeBaron}
\affiliation{University of California Santa Barbara \\
Santa Barbara, CA 93106, USA}
\author{Luca Angeleri}
\affiliation{University of California Santa Barbara \\
Santa Barbara, CA 93106, USA}
\author[0000-0002-6836-181X]{Samuel Whitebook}
\affiliation{University of California Santa Barbara \\
Santa Barbara, CA 93106, USA}
\author{Rachel Darlinger}
\affiliation{University of California Santa Barbara \\
Santa Barbara, CA 93106, USA}
\author{Phillip Morgan}
\affiliation{University of Colorado Boulder \\
Boulder, CO 80309, USA}
\author{Varun Iyer}
\affiliation{University of California Santa Barbara \\
Santa Barbara, CA 93106, USA}
\author[0009-0008-0190-3179]{Prerana Kottapalli}
\affiliation{University of California Santa Barbara \\
Santa Barbara, CA 93106, USA}
\author{Enda Mao}
\affiliation{University of California Santa Barbara \\
Santa Barbara, CA 93106, USA}
\author{Jasper Webb}
\affiliation{University of California Santa Barbara \\
Santa Barbara, CA 93106, USA}
\author{Dharv Patel}
\affiliation{University of California Santa Barbara \\
Santa Barbara, CA 93106, USA}
\author{Kyle Lam}
\affiliation{University of California Santa Barbara \\
Santa Barbara, CA 93106, USA}
\author{Kelvin Yip}
\affiliation{University of California Santa Barbara \\
Santa Barbara, CA 93106, USA}
\author{Michael McDonald}
\affiliation{University of California Santa Barbara \\
Santa Barbara, CA 93106, USA}
\author{Robby Odum}
\affiliation{University of California Santa Barbara \\
Santa Barbara, CA 93106, USA}
\author{Cole Slenkovich}
\affiliation{University of California Santa Barbara \\
Santa Barbara, CA 93106, USA}
\author{Yael Brynjegard-Bialik}
\affiliation{University of California Santa Barbara \\
Santa Barbara, CA 93106, USA}
\author{Nicole Efstathiu}
\affiliation{University of California Santa Barbara \\
Santa Barbara, CA 93106, USA}
\author{Joshua Perkins}
\affiliation{University of California Santa Barbara \\
Santa Barbara, CA 93106, USA}
\author{Ryan Kuo}
\affiliation{University of California Santa Barbara \\
Santa Barbara, CA 93106, USA}
\author{Audrey O'Malley}
\affiliation{University of California Santa Barbara \\
Santa Barbara, CA 93106, USA}
\author{Alec Wang}
\affiliation{University of California Santa Barbara \\
Santa Barbara, CA 93106, USA}
\author{Ben Fogiel}
\affiliation{University of California Santa Barbara \\
Santa Barbara, CA 93106, USA}
\author{Sam Salters}
\affiliation{University of California Santa Barbara \\
Santa Barbara, CA 93106, USA}
\author{Marlon Munoz}
\affiliation{University of California Santa Barbara \\
Santa Barbara, CA 93106, USA}
\author{Ruiyang Wang}
\affiliation{University of California Santa Barbara \\
Santa Barbara, CA 93106, USA}
\author{Natalie Kim}
\affiliation{University of California Santa Barbara \\
Santa Barbara, CA 93106, USA}
\author{Lee Fowler}
\affiliation{University of California Santa Barbara \\
Santa Barbara, CA 93106, USA}
\author{Philip Lubin}
\affiliation{University of California Santa Barbara \\
Santa Barbara, CA 93106, USA}

\begin{abstract}
\noindent We present the TRansient Image Processing Pipeline (TRIPP), a transient and variable source detection pipeline that employs both difference imaging and light curve analysis techniques for astronomical data. Additionally, we demonstrate TRIPP's rapid analysis capability by detecting transient candidates in near-real time.
TRIPP was tested using image data of the supernova SN2023ixf and from the Local Galactic Transient Survey (LGTS; \cite{LGTS}) collected by the Las Cumbres Observatory's (LCO) network of 0.4 m telescopes.
To verify the methods employed by TRIPP, we compare our results to published findings on the photometry of SN2023ixf.
Additionally, we report the ability of TRIPP to detect transient signals from optical Search for Extra Terrestrial Intelligence (SETI) sources. 

\end{abstract}

\keywords{Astronomy Data Analysis, SETI, Search for Extra Terrestrial Intelligence}

\section{Introduction} \label{Introduction}
\noindent Modern large-scale sky observing facilities and surveys such as the All-Sky Automated Survey for SuperNovae (ASAS-SN; \cite{Pojmanski2002}), Zwicky Transient Facility (ZTF; \cite{Bellm2014}), the Panoramic Survey Telescope and Rapid Response System 1 (PanSTARRS1; \cite{Chambers2016}), and the Asteroid Terrestrial-impact Last Alert System (ATLAS; \cite{Tonry2018}) have expanded survey data size into the petabytes. 
With the upcoming Vera Rubin Observatory Legacy Survey of Space and Time (LSST; \cite{LSSTScienceCollaboration2009}) imaging 5 petabytes of data annually, the size of survey data will only continue to grow.

The unprecedented influx of astrophysical data from modern sky surveys necessitates new data pipelines that can handle ever-increasing survey sizes. 
As the time to process and analyze an image often exceeds the speed of collecting images, surplus data is often stored in databases for asynchronous processing.
This strategy works poorly for time-sensitive transient phenomena like supernovae, fast radio bursts, solar system objects, gravitational wave sources for multi-messenger analysis, and other sources that change location or brightness on human or faster timescales.
For these time-sensitive phenomena, follow-up observations may not be possible if data processing is not expedient.

A quick, accurate, and multi-faceted method of processing the data generated by these endeavors is required to produce useful results and catalog important phenomena as they occur--in real time. To this aim, we present the TRansient Image Processing Pipeline (TRIPP), which employs difference imaging and light curve analysis techniques to identify transient candidates in near-real time.

\begin{figure}[hbt!]
    \centering
    \includegraphics[width = \columnwidth]{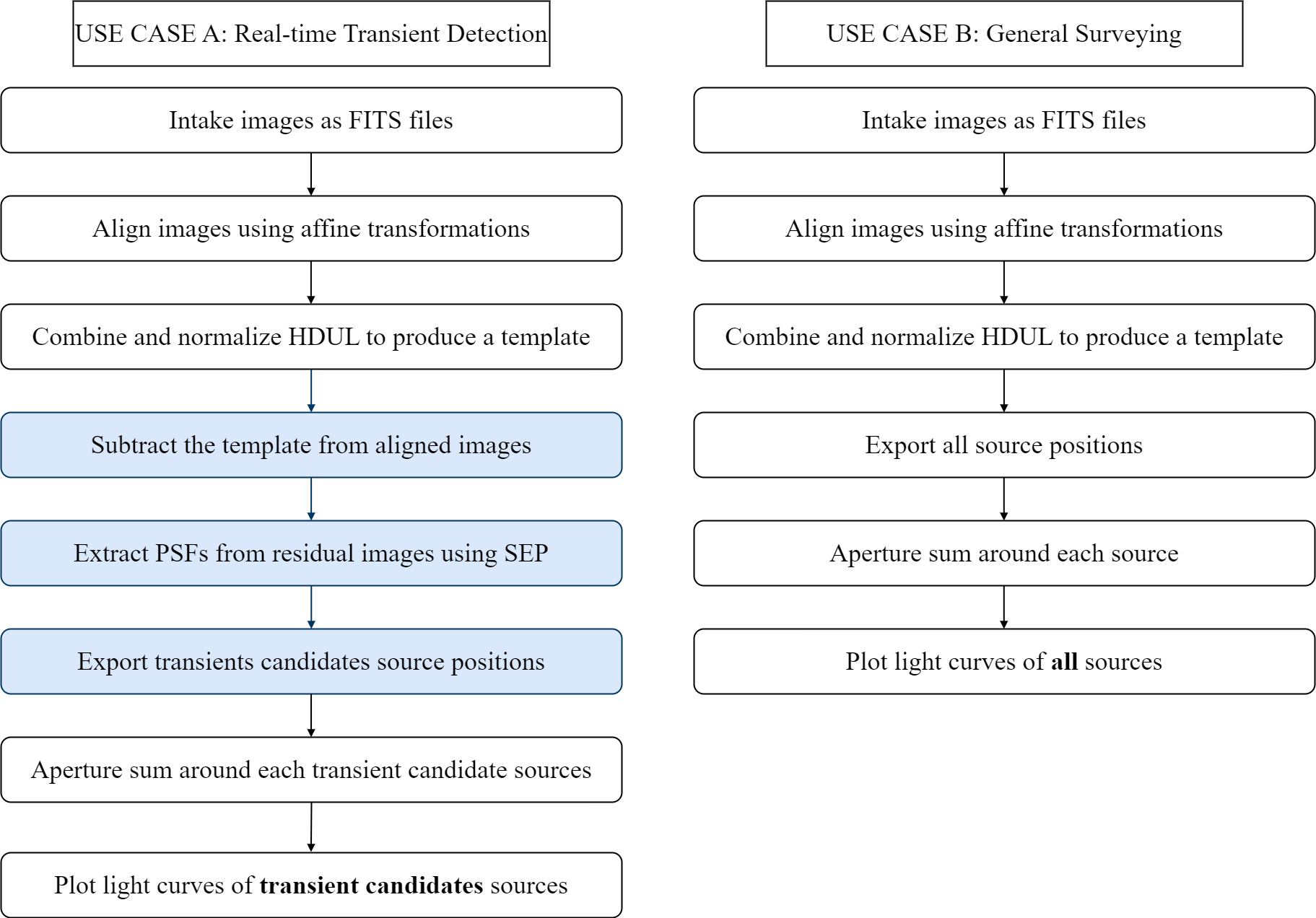}
    \caption{Flowchart of TRIPP. Preprocessed images are input as FITS files. They are aligned using Astroalign to a reference image that is specified by the user or automatically selected based on SNR, then combined into a median averaged template image. Performing DIA (shown in blue) can reduce the computation requirements for real-time transient detection by determining transient candidates and then only generating light curves for transient candidates.}
    \label{TRIPP_Flowchart}
\end{figure}

Difference imaging analysis (DIA) is a direct comparison method that can detect and measure transient variability. While there are many common techniques available (\cite{Alard1998}, \cite{Bramich2008}, \cite{Zackay_2016}), DIA is most simply explained as measuring the difference in optical brightness of a source between images taken at different times. 
In Figure \ref{TRIPP_Flowchart}, we propose using DIA to reduce the computational requirements for real-time analysis by only generating light curves of transient candidates, rather than all sources. Since successfully classifying a transient candidate in a single observing session requires significant variability during the observing session, all stellar sources should undergo asynchronous light curve analysis across many observing sessions to detect variability with longer characteristic timescales.

TRIPP begins by aligning preprocessed images and generating a median average value template image. For difference (residual) image analysis, TRIPP subtracts images from the template using a kernel subtraction method like Bramich and extracts transient candidates. For general surveying where we recommend TRIPP does not perform DIA (see Figure \ref{TRIPP_Flowchart}), TRIPP still generates a template image for reliable source extraction as the template averages out atmospheric effects refining the source location. For a space-based or adaptive optics instrument, that declines to use DIA, generating a template image for reliable extraction is likely unnecessary. 

Light curves can be produced for all sources present in a set of images or a list of candidate sources provided by the preceding pipeline steps.
Magnitudes of non-variable stars provided by the Sloan Digital Sky Survey's Data Release 17 (SDSS DR17; \cite{SDSS_DR17}) are used to perform photometric calibration of our data.
A variety of options are available for displaying calibrated light curves including phase folded Lomb-Scargle periodograms, systematic trend reduction using a median light curve, and sequential plotting by date. The final output is a CSV file containing all light curve data. In Figure \ref{TRIPP_Flowchart}, we give an overview of our processing procedures in the flowchart.

{}

We explain the data used to test TRIPP from our Local Galactic Transient Survey (LGTS) in Section \ref{data}, the TRIPP pipeline in Section \ref{pipeline}, and our results in Section \ref{results}.
In Section \ref{FotP}, we discuss our ongoing efforts with and preliminary results utilizing a convolutional neural network (You Only Look Once Darknet; \cite{Bochkovskiy2020}) for source extraction.
We end with an overview of our findings in Section \ref{conclusion}.

\section{Data} \label{data}
\subsection{LGTS Data and Optical SETI}
\noindent The majority of real data used to test TRIPP is from our Local Galactic Transient Survey (LGTS) using Las Cumbres Observatory Global Telescope network (LCOGT) 0.4 m telescopes to gather 6 megapixel images of the Andromeda Galaxy and the Magellanic Clouds.
In \cite{LGTS}, we explore the results of \cite{Lubin2016} to demonstrate how our general-purpose transient pipeline can utilize LGTS images for the Search for ExtraTerrestrial Intelligence (SETI). 
Based on our understanding of terrestrial technologies and future capabilities, SETI has historically focused on radio frequencies, which were well understood as means of communication when efforts began in the 1960s (see \cite{SETI}).
However, recent advances in laser technology suggest that lasers may be an effective means of interstellar communication.
Since few SETI searches have examined the optical wavelengths available with LCOGT telescopes, this is a natural extension of SETI efforts (\cite{Price2020}, \cite{Hippke2018}, \cite{Lubin2016}).
Our findings in \cite{LGTS} demonstrated that TRIPP could feasibly detect optical SETI signals in LGTS data with LCOGT 0.4m telescopes.
Therefore, by processing LGTS data, we aim to validate TRIPP's performance at relatively low signal-to-noise (due to the surface brightness of target galaxies) while making a concerted effort towards SETI.

Light curve analysis and DIA continue to be applied in recent surveys identifying naturally occurring transient sources, such as variable stars and supernovae discussed in \cite{Richmond2019}, \cite{Bonanos2019}, \cite{Moretti2018}, \cite{Morganson2018}, and \cite{Jencson2019}. To this end, we constructed our general-purpose TRansient Image Processing Pipeline (TRIPP) with these analysis methods. In addition to benefiting the wider Time-domain Astrophysics community, we hope TRIPP will benefit SETI through its capacity to identify technosignals by their proposed time domain nature.

\begin{figure}[hbt!]
    \centering
    \includegraphics[width = \columnwidth]{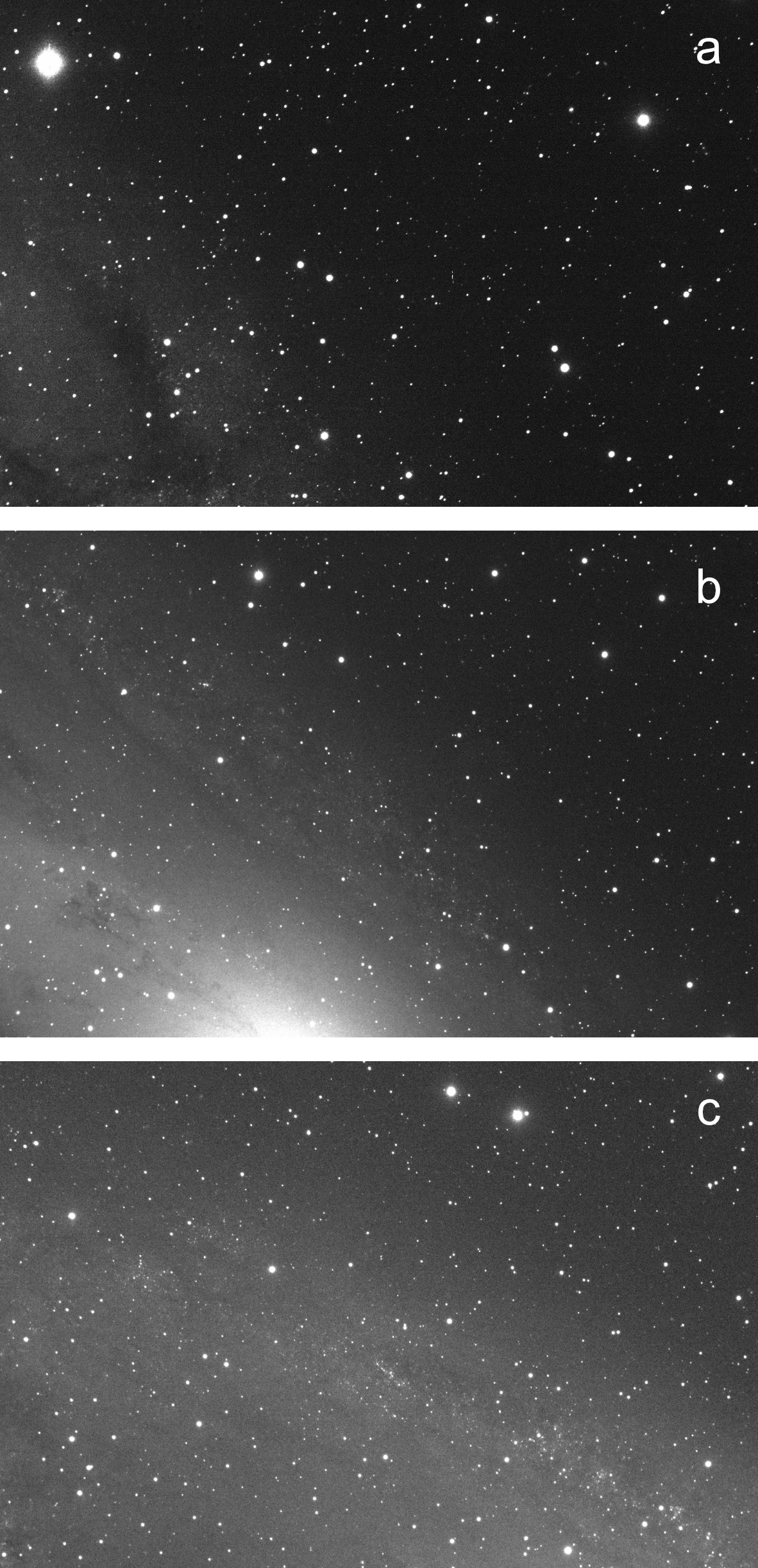}
    \caption{Typical starfields for TRIPP with 100 second exposure time for visibility. Images are BANZAI preprocessed. a) LGTS Section 38 taken at LCOGT's McDonald Observatory on 2019 October 18 using an integration time of 60 seconds. b) LGTS Section 23 taken at LCOGT's McDonald Observatory on 2019 July 19. c) LGTS Section 24 of M31 taken at LCOGT's Haleakala Observatory on 2019 August 18.}
    \label{images}
    \makeatletter\def\@currentlabel{7a}\makeatother\label{s38} 
    \makeatletter\def\@currentlabel{7b}\makeatother\label{s23} 
    \makeatletter\def\@currentlabel{7c}\makeatother\label{s24}
\end{figure}{}

\begin{figure}[hbt!]
    \centering
    \includegraphics[width = \columnwidth]{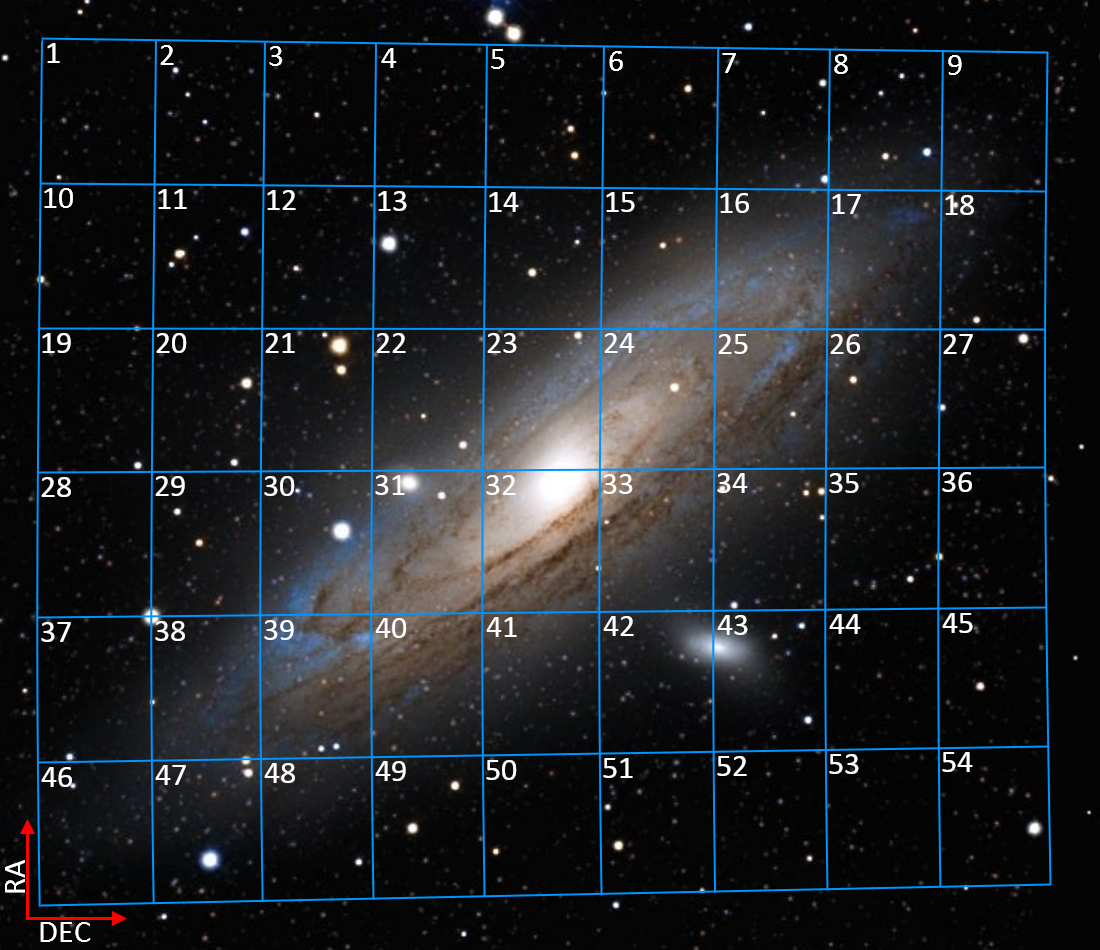}
    \caption{Andromeda LGTS survey sections. The sections of M31 imaged by LGTS prioritize galactic structure. Galactic image generated in Stellarium using a small FOV Mercator projection.}
    \label{M31}
\end{figure}

Since the angular size of Andromeda exceeds the field of view of LCOGT's 0.4 m telescopes (29.2$'$ x 19.5$'$), LGTS observed Andromeda in sections as shown in Figure \ref{M31}. Only sections that contain Andromeda were collected to test the pipeline's capabilities against high surface brightness.
Some typical images from TRIPP can be seen in Figure \ref{images} showing the varied background magnitude of Andromeda as a function of radial distance from the galactic center.
The 2023 upgrade of LCO's 0.4 m network enables wide-field imaging with a vastly improved 1.9 x 1.3 degree field. 
This upgrade will greatly expedite future surveying by requiring only three sections to survey the entirety of the Andromeda galaxy \cite{Harbeck2023}.

{}

\subsection{SN2023ixf Data}
\begin{figure}[hbt!]
    \centering
    \includegraphics[width = \columnwidth]{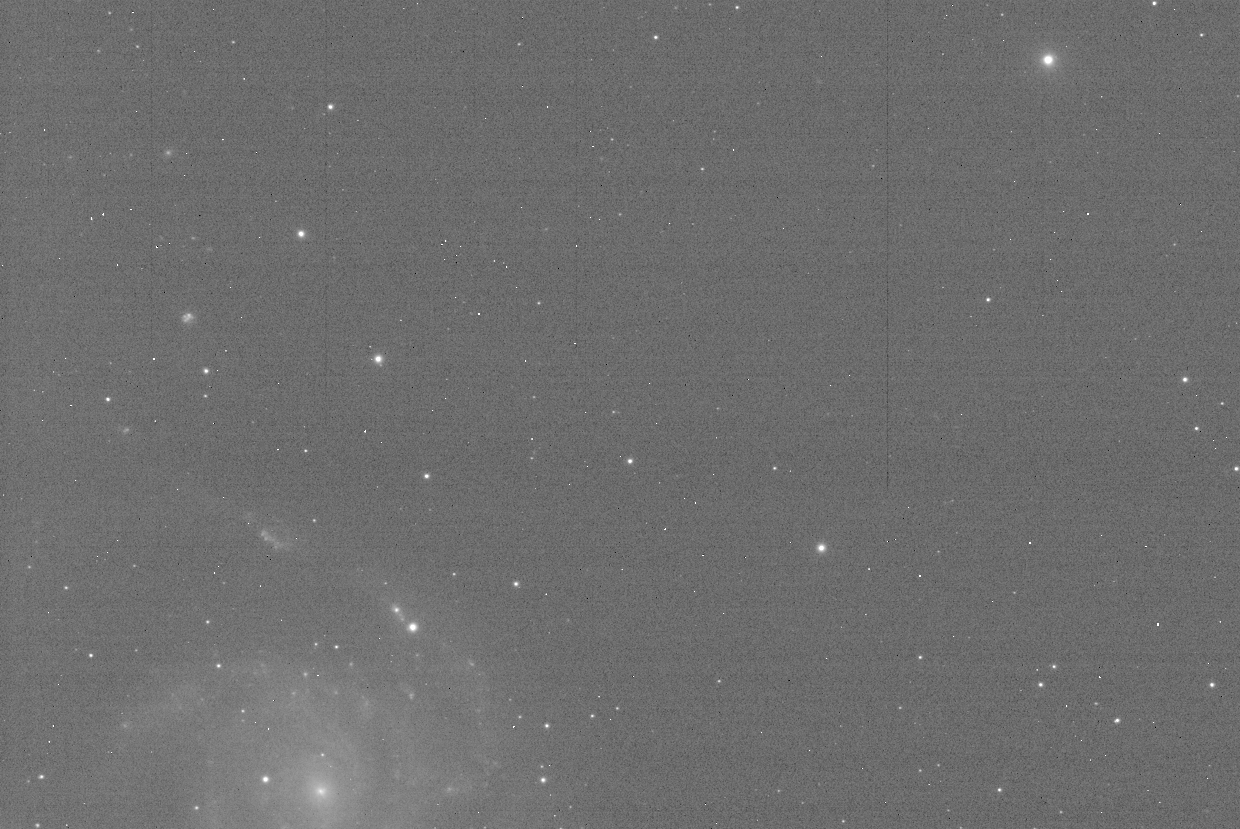}
    \caption{Median stacked image of the twenty-five 15 second exposures from our first observing night. This 29.2$'$ x 19.5$'$ field of view contains SN2023ixf, a portion of M101, and SDSS cataloged reference stars.}
    \label{sntemp}
\end{figure}

\noindent We also observed the nearby supernova SN2023ixf to validate TRIPP's photometry process and results.
Our data for SN2023ixf spans May 25th through July 11th of 2023, again collected by the LCOGT Network's array of 0.4 m telescopes.
Out of twelve total observing attempts, ten produced usable data with two nights excluded due to poor seeing conditions.
Each observing run consists of twenty-five, 15 second exposures in SDSS-r' and SDSS-i' bands.
As discussed later in Section \ref{cal}, TRIPP relies on SDSS reference magnitudes to calibrate light curves.
As a result, we did not choose SN2023ixf as the telescope's pointing target.
Instead, we chose pointing coordinates that would include most of M101, the supernova, and a sufficient number of SDSS cataloged reference stars.
This field of view can be seen in Figure \ref{sntemp}. 

\section{\footnotesize Transient Image Processing Pipeline (TRIPP)} \label{pipeline}
\subsection{Image Alignment}
We perform positional image alignment with affine transformations using the OpenCV library. A set of three points is selected in a reference image. The pixel values of these three points are converted into RA and DEC using the WCS information provided by the FITS header. This set of RA and DEC coordinates is then transformed back into pixel values in the images to be aligned. The OpenCV function \texttt{getAffineTransform()} is used to generate the alignment transformation between images using the selected target points and applied with the \texttt{warpAffine()} function. This method provides an accurate alignment method while boasting a short run time.

\begin{figure}[hbt!]
    \centering
    \includegraphics[width = \columnwidth]{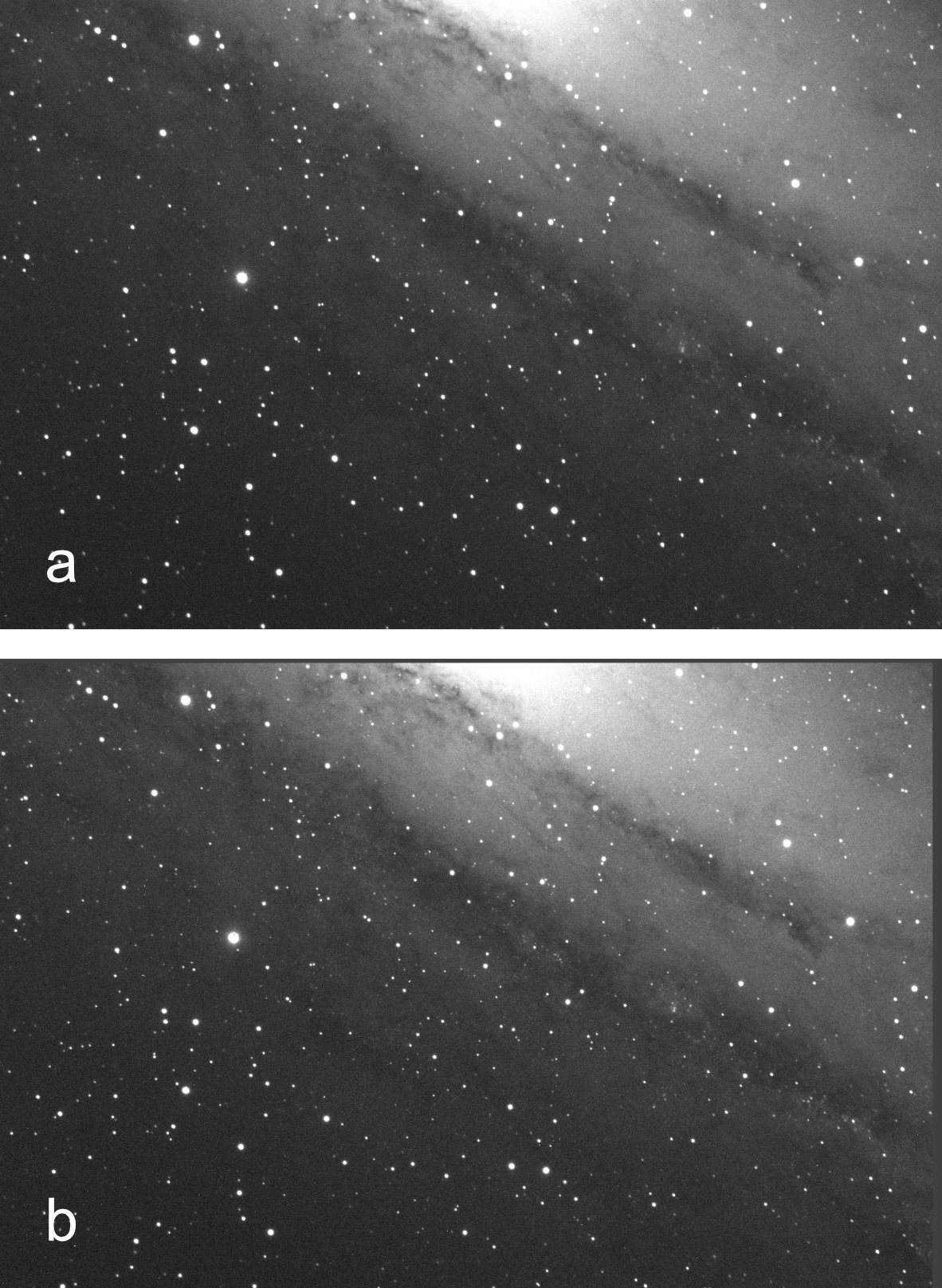}
    \caption{Single 10 second exposure LGTS image of M31 section 41 that has only gone through preprocessing (a) versus the same image after the align step (b). The lines along the edges (b) are an artifact of the alignment process which the extract step excludes due to their size.}
    \label{aligned}
\end{figure}{}

\subsection{Difference Imaging} \label{section:diffimaging}
\begin{figure}[hbt!]
    \centering
    \includegraphics[width = \columnwidth]{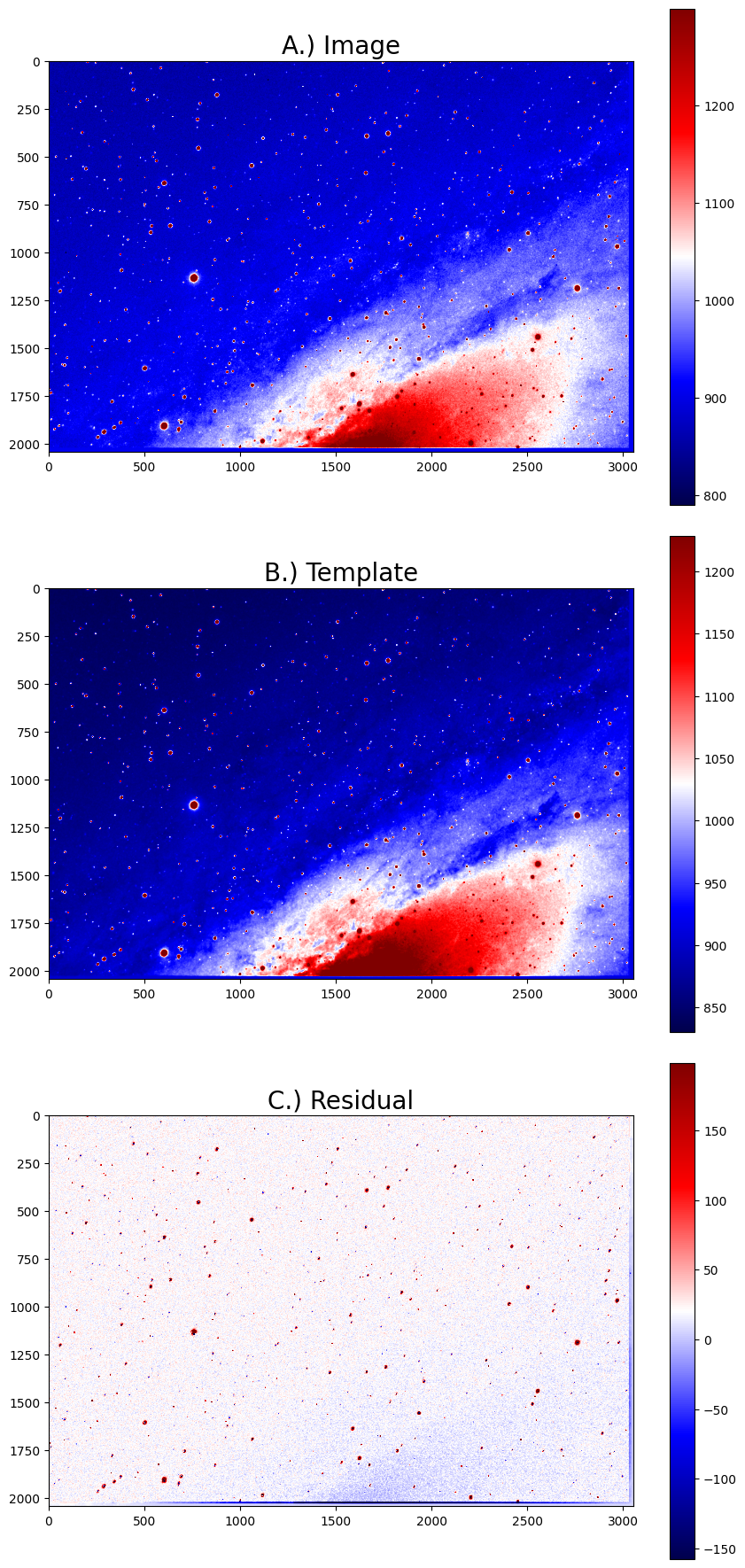}
    \caption{10 second exposure LGTS images of M31 section 41 which showcase TRIPP's image subtraction in a dense star field. Image A is a single image of the section from which the combined template Image B is subtracted via SFFT to achieve the residual Image C. In this image, the background average was 86.46 ADU with 569 sources above the 5$\mu_{bkg}$ threshold. A common overscan artifact is present along the lower and righthand sides of the images and rejected by extract due to its size.}
    \label{41subtract}
    \makeatletter\def\@currentlabel{6a}\makeatother\label{41subtracta} 
    \makeatletter\def\@currentlabel{6b}\makeatother\label{41subtractb}
    \makeatletter\def\@currentlabel{6c}\makeatother\label{41subtractc} 
\end{figure}

\begin{figure}[hbt!]
    \centering
    \includegraphics[width = \columnwidth]{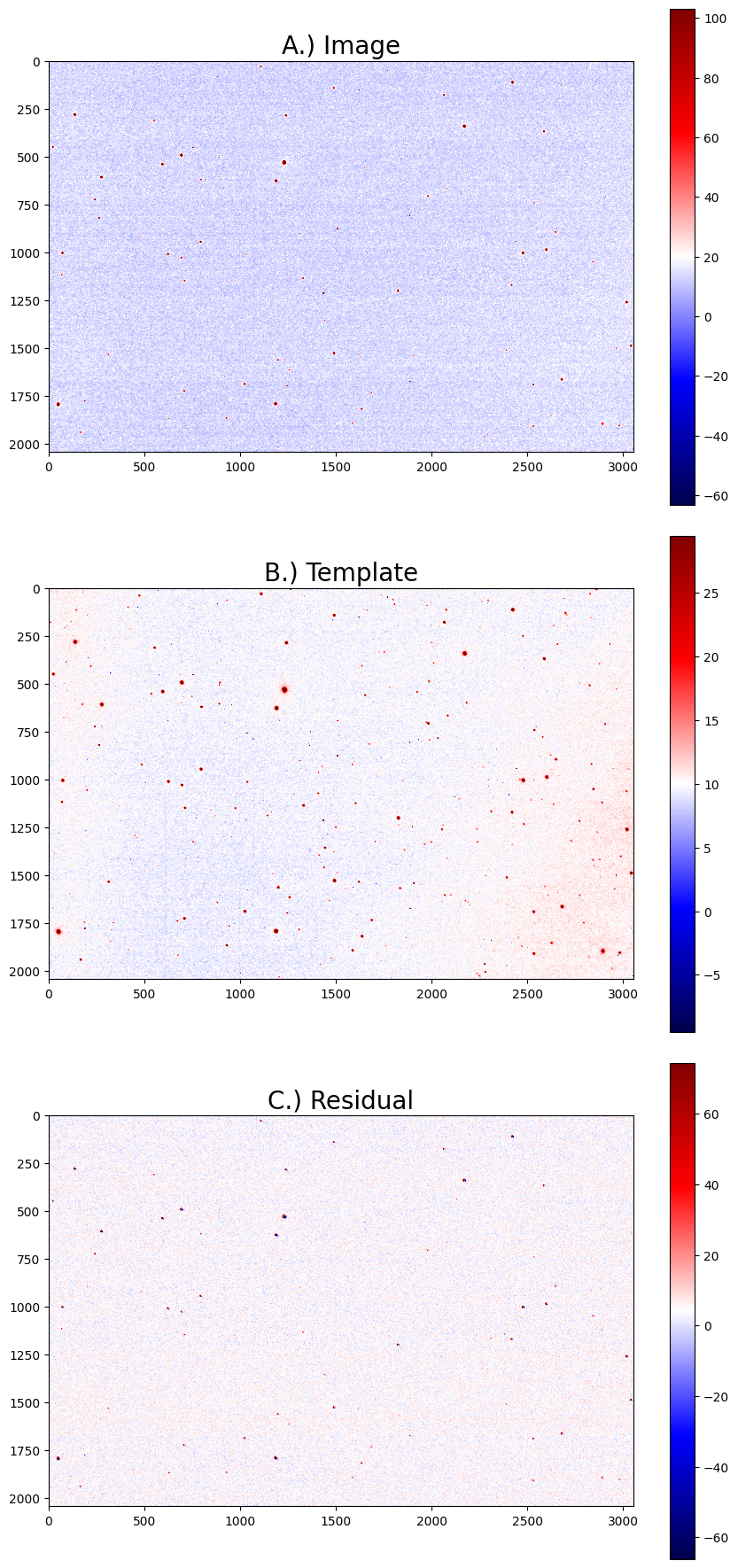}
    \caption{10 second exposure LGTS images of M31 section 24 which showcase TRIPP's image subtraction in a dense star field. In this image, the background average was 43.5 ADU with 64 sources above the 5$\mu_{bkg}$ threshold.}
    \label{24subtract}
    \makeatletter\def\@currentlabel{7a}\makeatother\label{24subtracta} 
    \makeatletter\def\@currentlabel{7b}\makeatother\label{24subtractb}
    \makeatletter\def\@currentlabel{7c}\makeatother\label{24subtractc} 
\end{figure}

\noindent After each image in the input dataset is aligned, the images are combined into a median value template image. This process then takes the median value at each pixel location across the set of images to produce a template.
Following \cite{Drake2009}, this template image is used for subtraction with examples shown in Figures \ref{41subtract} and \ref{24subtract}. 

To create difference (residual) images, we subtract the template image created in the image combination step from each image in the input dataset.
We do this because if any transient sources are present, subtracting their median value will make variability quantitatively apparent.
TRIPP currently utilizes one of four different subtraction methods depending on user choice.
The comparison offered in this section was computed with Dual Intel Xeon Gold 6154s, an NVIDIA RTX A6000 GPU (48 GB of VRAM), and 512 GB of DDR4 RAM on Ubuntu 18.04.

The most resource efficient method is a direct numerical image subtraction where each image is represented as a matrix of pixel values which are then subtracted from each other using numeric python (NumPy).
However, each source in an image is represented as a point spread function (PSF) and must be aligned, which is separate from image alignment. 
For this reason, a simple matrix subtraction of two images often yields artifacts from sources in the original image as a result of misalignments in the PSF peaks between the two images which other subtraction methods aim to remove. 

PSF misalignments are mitigated by algorithms such as Bramich which employ convolution kernels to solve the PSF alignment issue by least squares regression (\cite{Miller2008}) as seen in Figure \ref{41subtract}.
Bramich has a variant, Adaptive Bramich, which allows a variable prefactor for space-varying delta convolution kernels to better align PSFs between images, resulting in improved performance in dense star fields.
The Bramich methods are implemented using the Optimal Image Subtraction (OIS; \cite{OIS}) module for Python.
The drawback of Adaptive Bramich is computation time.
Each convolution kernel associated with a given PSF has a large number of fitting parameters (quadratic with the degree of the polynomial used to model the differential background variation) that need to be minimized.
As a result, this method exceeds one minute per image using GPU acceleration on an RTX A6000.
Without GPU acceleration, Adaptive Bramich takes 40 minutes per image on our hardware.

On the other hand, the regular Bramich method with the static convolution kernel has slightly worse performance in dense star fields.
However, it comes with a significant computation time boost; coming in at just 30 seconds per image.
Moreover, we dissected this algorithm and added a host of hardware and software optimizations like offloading heavy compute tasks to the GPU, replacing for loops with NumPy arrays for linear algebra calculations, and reusing identical variables across images over recalculating them each time.
These optimizations reduced the total processing time by greater than 100x and our subtraction times down to hundreds of milliseconds.
Since our optimized Bramich is now much faster and acceptable for most conditions, it has become the primary subtraction algorithm utilized for LGTS data and we recommend it as the default subtraction method for most users.

We have also implemented the Saccadic Fast Fourier Transform (SFFT; \cite{Hu_2022}) method which uses a delta-function basis for kernel decomposition and performs image subtraction in Fourier Space.
The kernel and background polynomials can also be specified within SFFT.
TRIPP supports this flexibility and defaults to a linear kernel and background polynomial which was used in all SFFT figures and timings presented.
As this method offers a substantial computational time improvement over Adaptive Bramich and native GPU utilization, SFFT is optimal for users who are seeking to implement real-time DIA but require a variable kernel subtraction solution.

\subsection{Transient Source Extraction} \label{extraction}
\noindent Transient source candidates are identified within residual images using Source Extractor (\cite{Bertin1996}).
For faster processing, we use Source Extractor Python (SEP; \cite{Barbary2016}) which accelerates the module.
On linear detectors, the value measured at each pixel is the sum of background noise and signal from observed sources.
In order to extract sources, SEP first estimates the background noise of each residual frame.
This estimated background is then subtracted from the residual and followed by source extraction.
The threshold for source extraction in TRIPP is configurable based on the root mean square of the background in the subtracted image $\mu_\textrm{bkg}$ and was set to $5 \mu_\textrm{bkg}$ in LGTS testing.
We also reject all PSFs that contain more than 250 pixels to automatically remove artificial transients created by alignment such as the example in Figure \ref{41subtractc}.

\subsection{Photometry} \label{Aperture Phot}
\noindent Photometric light curves are generated using aperture photometry, calibrated using data from the Sloan Digital Sky Survey's Data Release 17 (SDSS DR17; \cite{SDSS_DR17}), and output at the end of TRIPP's operation.
SDSS was chosen as our reference catalog as our data is captured using SDSS filters, in addition to offering extensive photometric coverage of the night sky.

Photometry in TRIPP borrows from the early steps of the pipeline, completing the align and combine steps before branching away from the subtraction step as shown in \ref{TRIPP_Flowchart}.
Sources are identified from the template image using SEP, as outlined in Section \ref{extraction}.
If a user inputs multiple observing sessions, a template is created for each session.
This allows sources that only appear due to inconsistencies in telescope pointing to be excluded.
This initial list of sources is then queried through the SDSS database using the Astroquery Python package.
Source positions in RA and DEC are calculated from pixel position using the WCS transformation provided by LCO's BANZAI preprocessing.
These positions are then used in a cone search with a 15 arcsec radius to obtain photometric data.
If the search returns photometric data, TRIPP utilizes SDSS's \texttt{psfMag} and \texttt{psfMagERR} methods to calculate source magnitudes and magnitude errors.
TRIPP obtains the observation filter from the FITS header in order to query the apparent magnitude in the appropriate passband. SDSS encourages the use of its `clean' photometry search parameter which flags `bad' photometry.
However, we found that the `clean' photometry search parameter was often too aggressive in excluding sources for our needs. Instead, we allow all sources that are designated as stars and brighter than a user defined magnitude threshold (typically magnitude 15) to be reference sources with the uncertainty in SDSS photometry accounted for in the calibration step. 

Aperture photometry is then performed for each source.
SEP provides positions and bounding pixel coordinates, $x_{max}$ and $x_{min}$ for each extracted source.
These parameters determine the location and size of the aperture within an image. Background subtraction is performed locally for each source using a background annulus.
For all sources, the `background annulus' inner radius, $r_{ann_i}$ is defined as being five pixels larger than the source aperture's radius.
The outer radius of the annulus, $r_{ann_o}$ is a further five pixels larger than the inner radius. 
\begin{equation}
    r_{ap} = \frac{(x_{max} - x_{min})}{2} 
\end{equation}
\begin{equation}
    r_{ann_i} = r_{ap} + 5 
\end{equation}
\begin{equation}
    r_{ann_o} = r_{ann_i} + 5 = r_{ap} + 10
\end{equation}
After BANZAI's normalization and dark field subtraction, LGTS images have a reported gain of 1.0 and dark current of 0.00.
This allows us to sum the pixel values, $N(j, k)$ within a given aperture to obtain a raw flux value.
Local background subtraction is performed by subtracting the summed pixel values in the outer annulus $N(u, w)$, scaled by the ratio of the number of pixels in the aperture and annulus, $N_{ap}$ and $N_{ann}$, respectively.
This method aims to account for light spreading across multiple pixels.
The total flux, $n$, for a source in units of $e^{-}$ is then:
\begin{equation}
    n = \sum_{j, k} N(j,k) - (\frac{N_{ap}}{N_{ann}})* \sum_{u, w} N(u, w)
\end{equation}
In some cases, the background subtraction can yield a negative flux value when a source is dim.
These sources are flagged and excluded from further calculations.
The final flux value, $n$, is then used to calculate an instrumental magnitude:
\begin{equation}
    m_{i} = -2.5 * log(n)
\end{equation}
A statistical error for each instrumental magnitude measurement is then calculated as follows.
 The error in the background subtracted flux is: 
\begin{equation}
    \delta n = \langle \delta n^{2} \rangle^{\frac{1}{2}} = \left[\sum_{j,k} (\delta N^{2}) + (\frac{N_{ap}}{N_{ann}})^{2} * \sum_{u,w} (\delta N^{2})\right]^{\frac{1}{2}}
\end{equation}
Where $\delta N$ is the uncertainty in a given pixel measurement given by:
\begin{equation}
    \delta N = (N_{R}^{2} + (\sqrt{N})^2)^{\frac{1}{2}} = (N_{R}^{2} + N)^{\frac{1}{2}}
\end{equation}
 With $N_{R}$ as the readout noise from the detector for a given frame.
 This gives a total uncertainty in each flux measurement of: 
\begin{equation}
    \delta n = \left[\sum_{j,k} (N_{R}^{2} + N(j, k)) + (\frac{N_{ap}}{N_{ann}})^{2} * \sum_{u,w} (N_{R}^{2} + N(u,w))\right]^{\frac{1}{2}}
\end{equation}
The final error in the instrumental magnitude is then: 
\begin{equation}
    \label{final err}
    \delta m_{i} = 2.5*log(e)*\abs{\frac{\delta n}{n}}
\end{equation}

\subsection{Calibration} \label{cal}
\noindent TRIPP instrumental magnitudes are calibrated to a reported on-sky magnitude using reference stars in SDSS.
Instrumental magnitudes of sources in TRIPP are calculated for every image.
After this, a linear Orthogonal Distance Regression (ODR) fit is done between stars that exist in both the SDSS and TRIPP catalogs, are brighter than the magnitude threshold, and are not on the image border. 
SDSS photometric error is calculated using the same process as TRIPP.
As such, we assume an equal variance of error between SDSS and TRIPP ($\delta = 1$), and the calibration coefficients can be expressed directly in terms of the second degree statistical sample moments of reference sources.

Calibration coefficients are calculated per frame to minimize additional error injected from background brightness variation.
The on-sky magnitude for each frame, $M$, is then given as

\begin{equation}
    \label{eq: ZP}
    M = \beta_0 m_i + \beta_1
\end{equation}
where $m_i$ is the source instrumental magnitude and $\beta_n$ are calculated as derived in \cite{Glaister2001}, and y and x are the SDSS and TRIPP reference magnitudes respectively.
\begin{equation}
    \label{eq:beta_1}
    \beta_1 = \frac{s_{yy} - s_{xx} + \sqrt{(s_{yy} - s_{xx})^2 + 4s_{xy}^2}}{2 s_{xy}}
\end{equation}

\begin{equation}
    \label{eq:beta_0}
    \beta_0 = \Bar{y} - \beta_1 \Bar{x}
\end{equation}
Where $s_{ij}$ are the covariance terms.
A mean absolute error (MAE) calculation is then used to provide an overall error estimation for the calibration fit.
\subsection{Variable Source Identification}
\noindent To determine if a source is variable, TRIPP relies on a chi-squared per degree of freedom $\chi_{dof}^2$ metric for a weighted average magnitude defined by: 
\begin{equation}
    M_{avg}= \frac{\sum_{i = 1}^n w_{i}m_{i}}{\sum_{i = 1}^{n} w_{i}}
\end{equation}
Where $w_i$ is the weight associated with each magnitude measurement $m_i$ given by:
\begin{equation}
    w_i = \frac{1}{\delta m_{i}^2}
\end{equation}
the square reciprocal of Equation \ref{final err}. The $\chi_{dof}^2$ is then given by: 
\begin{equation}
    \chi^2 = \sum_i \frac{(m_i - M_{avg})^2}{\delta m_{i}^2}
\end{equation}
\begin{equation}
    dof = (n-1)
\end{equation}
Where $n$ is the total number of magnitude measurements, finally giving:  
\begin{equation}
    \chi_{dof}^2 = \frac{\chi^2}{dof}
\end{equation}

A source which is non-variable should have a $\chi_{dof}^2$ on the order of unity. However, we relax this assumption to avoid flagging every source with only minor statistical variations in magnitude.
Therefore, by default, TRIPP flags sources with a $\chi_{dof}^2$ of greater than 25 as variable sources.
This variability threshold can be changed by the user to better suit their needs.  

\section{Results} \label{results}
\subsection{Subtraction Capability}\label{subtractioncapability}
\noindent The residuals from images subtracted using Adaptive Bramich have a smaller standard deviation and absolute mean than all Bramich residuals, especially when there is a noticeable variation in the background luminosity.
When subtracting identical star fields with no transients, Bramich subtraction yielded an average standard deviation of 14.8 compared to an average standard deviation of 13 for Adaptive Bramich and 17.9 for SFFT with a mean of 0 for all.
Since Adaptive Bramich yields the least variance, Adaptive Bramich has the highest reliability at the expense of the slowest computation time.
Moreover, we present an algorithmic optimization that demonstrates superior computational efficiency compared to the previous implementation (\texttt{ois}).
Our changes achieved significantly faster execution times(${\sim}100$x, from ${\sim}30$ seconds to ${\sim}300$ milliseconds).
Furthermore, our performance analysis with traditional Bramich from \cite{OIS} did not indicate any difference between the output image quality.
By leveraging existing advanced computational hardware, namely GPUs and large cache sizes, our algorithm offers a notable reduction in processing time and has substantial implications for real-world applications of this image processing code-base.
These results emphasize the potential for our approach to enhance the performance of DIA, particularly in scenarios where time-sensitive results are important and storage space is finite.
The exact methodology needed to achieve this result is described in Section \ref{section:diffimaging} which will be available for general use with this pipeline on GitHub under the MIT license.

In section 24 which ranges from near the core to outside the galaxy (fig \ref{s24}), Bramich and Adaptive Bramich yielded similarly lower average deviations of 15.9 and 16.1, respectively, while SFFT yielded 19.5.
However, the average background in the subtracted image $\mu_\textrm{bkg}$ was considerably lower for SFFT than Bramich methods.
This difference is expected as SFFT utilizes Fourier Transforms to pick out any sources present in a local region and performs subtraction on those sources alone, whereas the convolution employed by Bramich is less precise.
Since SFFT can easily pick out the background noise due to the nature of Fourier Transforms, it can more precisely remove background noise.

Timing information for each subtraction method is displayed in Table \ref{table:timing}.
This table along with the subtraction accuracy discussed earlier in this section makes it clear why the ideal choice for LGTS data is the optimized version of Bramich. 

\begin{deluxetable}{lcccc}[htb]
\tablecolumns{5}
\tabletypesize{\scriptsize}
\setlength{\tabcolsep}{1pt}
\tablecaption{\label{table:timing} Timings of TRIPP per 6 megapixel LGTS image}
\tablehead{\colhead{DIA method} & \colhead{NumPy} & \colhead{Bramich}& \colhead{Adaptive Bramich} & \colhead{SFFT}}
\startdata
    Read                & $5.7\pm 0.2$ ms              & $5.4\pm 0.2$ ms                 & $5.4\pm 0.2$ ms                & $5.4\pm 0.2$ ms       \\ \hline
    Align               & $1.7\pm 0.9$ s               & $1.2\pm 0.6$ s                  & $1.2\pm 0.6$ s                       & $1.2\pm 0.6$ s  \\ \hline
    \textbf{Subtract}   & $\mathbf{0.12\pm 0.03}$ \textbf{s} & $\mathbf{0.26\pm 0.03}$ \textbf{s} & $\mathbf{70\pm 30}$ \textbf{s} & $\mathbf{4\pm1}$ \textbf{s} \\ \hline
    Extract             & $0.29\pm 0.09$ s              & $0.3\pm0.14$ s      & $0.32\pm0.14$ s                 & $0.3\pm0.1$ s               \\ \hline
    Write               & $0.32\pm 0.10$ s              & $0.38\pm 0.16$     & $0.38\pm 0.16$                 & $0.37\pm 0.11$ s            \\ \hline
    Average             & $2.6\pm0.8$ s                 & $2.19\pm 0.66$     & $74\pm4$ s                     & $6\pm0.6$ s
\enddata
\tablecomments{These results are averaged from images in sections 23, 32, 24, and 38 processed using 36 total cores, an RTX A6000 GPU (48 GB of VRAM), and 512 GB of DDR4 RAM. Processing time for each step is independent of the subtraction method. Average time does not include preprocessing (handled by LCO) or light curve creation which is proportional to the number of transient candidates.}
\vspace{-0.8cm} 
\end{deluxetable}

\subsection{Extraction Results}
\noindent To ascertain the abilities of Source Extractor to extract transient sources from processed images, we created simulated images using the \textsc{Stuff} and \textsc{SkyMaker} programs described in \cite{Bertin_2009}. To highlight the possibility of machine learning to identify transients in non-ideal conditions, we use a relatively bright background magnitude of 18 arcsec$^{-2}$. The transients range from magnitude 6 to 19 with variable magnitude differences. A correct detection, or true positive, was determined by reporting a transient location within 10 pixels of the true center given to the simulator.

To measure the prediction performance of the method we use a $F_1$ statistic defined as:
\begin{equation}
    F_1 = 2\frac{PR}{P+R}
\end{equation}
Across these simulated image pairs, we found an $F_1$ statistic for transient candidates of 0.27(0.38) utilizing a threshold of the source exceeding 3(5)$\mu_\text{bkg}$ in the residual image as the threshold for classification. However, this $F_1$ statistic is dominated by false positives caused by Source Extractor's inability to distinguish Bramich artifacts from transients. Due to the deconvolution solution in Fourier space, Bramich (along with many other forms of subtraction) produces artifacts and amplified noise (\cite{Zackay_2016}). To overcome this limitation, we are exploring machine learning extraction, as described in \ref{FotP}, which boasts a preliminary $F_1$ statistic of 0.72.

Following \cite{li2022automated}, we define the precision $P$, and recall $R$ of Source Extractor as

\begin{equation}
    P = \frac{TP}{TP + FP},
\end{equation}

\begin{equation}
    R = \frac{TP}{TP + FN}
\end{equation}

where $TP$ is the cumulative number of true positives, $FP$ is the cumulative number of false positives, and $FN$ is the cumulative number of false negatives.

The computer vision offered by ML extraction has been trained to identify and reject artifacts leading to hundreds of times fewer false positives and higher precision. However, ML's strictness also conversely decreases the number of true positives by a factor of ${\sim}3$ yielding lower recall. Therefore, machine learning extraction is considerably more accurate than comparing to the average background as shown by the improved $F_1$ statistic. A summary of these results is presented in Table \ref{table:extract}. This gain in $F_1$ statistic comes with the strong drawback of introducing failed detections. As identifying a false transient candidate only requires photometry for elimination, while a false negative could result in an undetected transient, future work in this area must focus on fine-tuning the ML model to enhance recall without significantly hindering precision.

\begin{deluxetable}{ccccccc}[hbt]
\tablecolumns{7}
\tabletypesize{\scriptsize}
\tablecaption{Comparison of traditional and machine learning transient extraction methods. \label{table:extract}}
\tablehead{
    \colhead{Extract} & \colhead{True} & \colhead{False} & \colhead{False} & \colhead{Precision} & \colhead{Recall} & \colhead{F1} \\[-0.1in]
    \colhead{Method} & \colhead{positives} & \colhead{positives} & \colhead{negatives} & \colhead{} & \colhead{} & \colhead{}
}
\startdata
    3$\mu_{\textrm{rms}}$   & 21383  & 115751    & 1264      & 0.16      & 0.94      & 0.27 \\
    5$\mu_{\textrm{rms}}$   & 20990  & 66752    & 1933      & 0.24      & 0.92      & 0.38 \\
    ML       &  6750 &      187     &     5161      &  0.97         &  0.57         & 0.72 \\
\enddata
\vspace{-0.8cm} 
\end{deluxetable}

\subsection{Light Curve Results}
\begin{figure}[hbt!]
    \centering
    \includegraphics[width = \columnwidth]{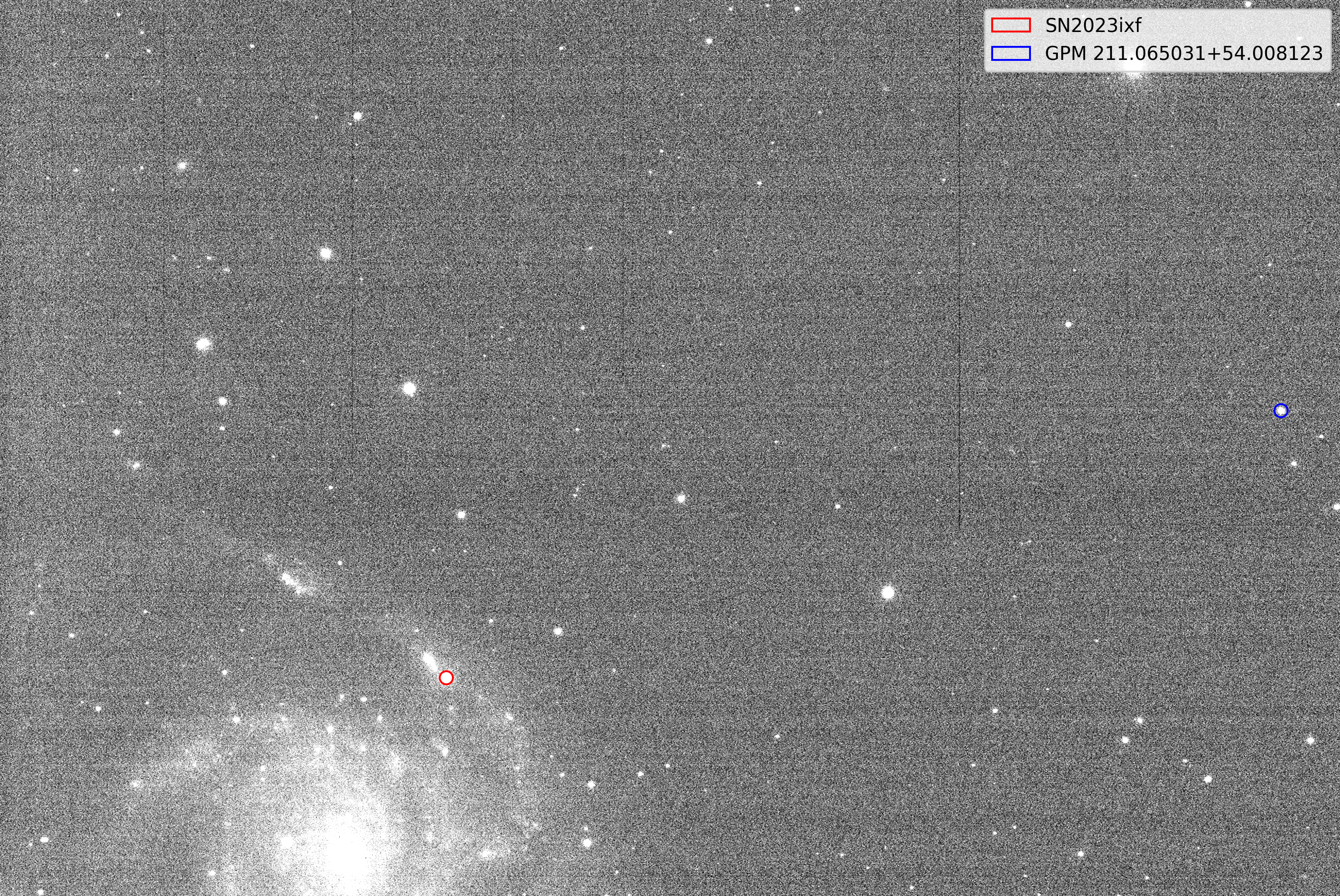}
    \caption{The locations of GPM 211.065031 +54.008123 and SN2023ixf identified in this field.}
    \label{source_locations}
\end{figure}

\noindent Our field of view containing SN2023ixf was 29.2$'$ x 19.5$'$, allowing us to capture numerous non-variable sources in addition to SN2023ixf. We report a light curve of SN2023ixf along with an example light curve of a non-variable source as validation of our photometry procedures. Their positions are circled and displayed in Figure \ref{source_locations}.

\begin{figure}[hbt!]
    \centering
    \includegraphics[width = \columnwidth]{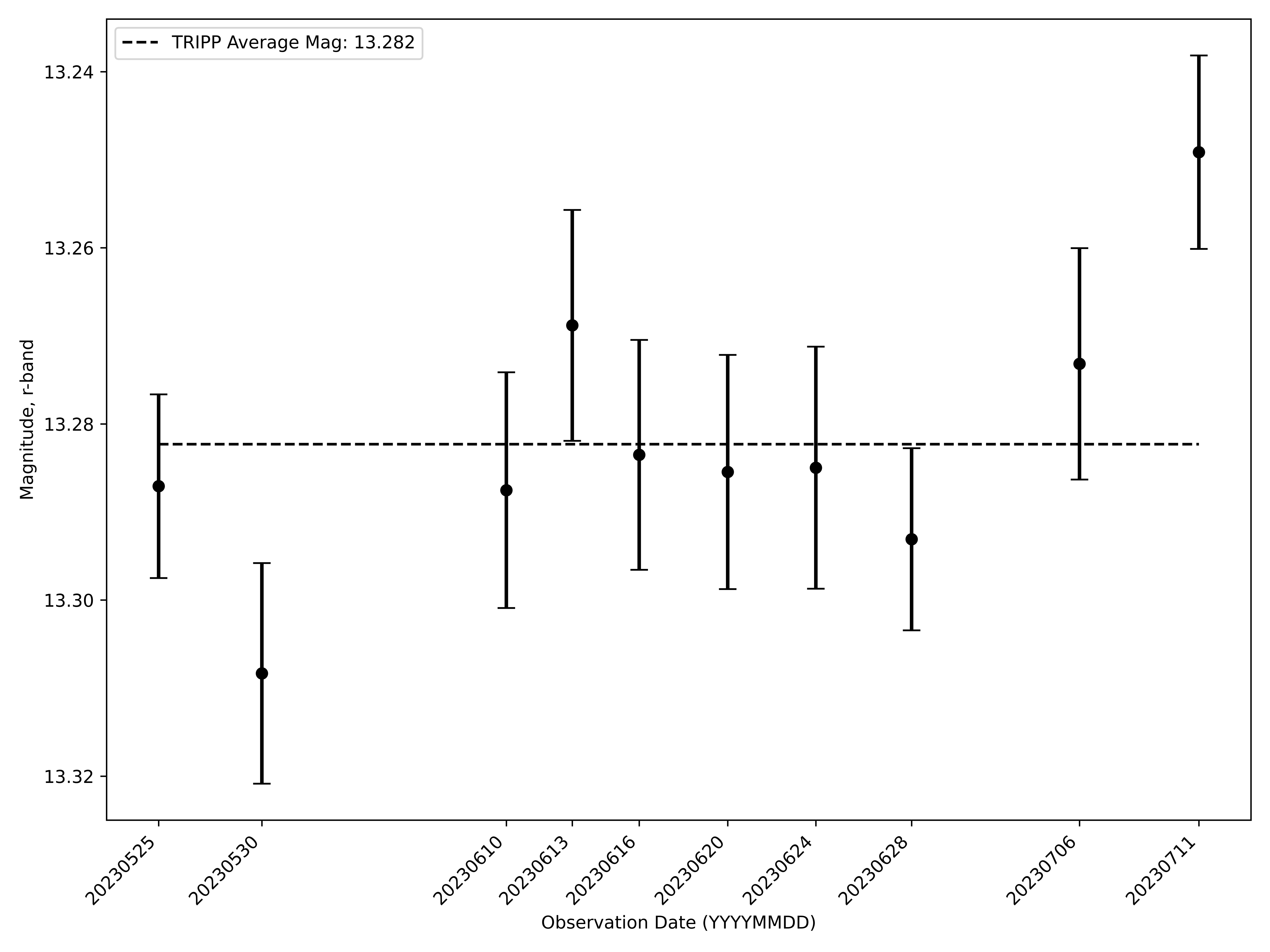}
    \caption{Light curve for GPM 211.065031+54.008123 spanning ten observing nights in SDSS-r' band. Each observation consists of twenty-five 15 second exposures. The short integration time was selected to avoid saturation in SN2023ixf.}
    \label{source58 light curve}
\end{figure}

The $\chi_{dof}^2$ for GPM 211.065031+54.008123 \ref{source58 light curve} compared to a non-variable hypothesis using the weighted average magnitude is 2.23.
Our calculated magnitudes are typically constrained to a few tenths of a magnitude and agree with statistical errors.
In this case, our magnitudes only differ by less than 5 millimag.
Magnitudes off by more than that can be explained by ``unclean'' photometry from SDSS as described in Section \ref{Aperture Phot}.

\begin{figure}[hbt!]
    \centering
    \includegraphics[width = \columnwidth]{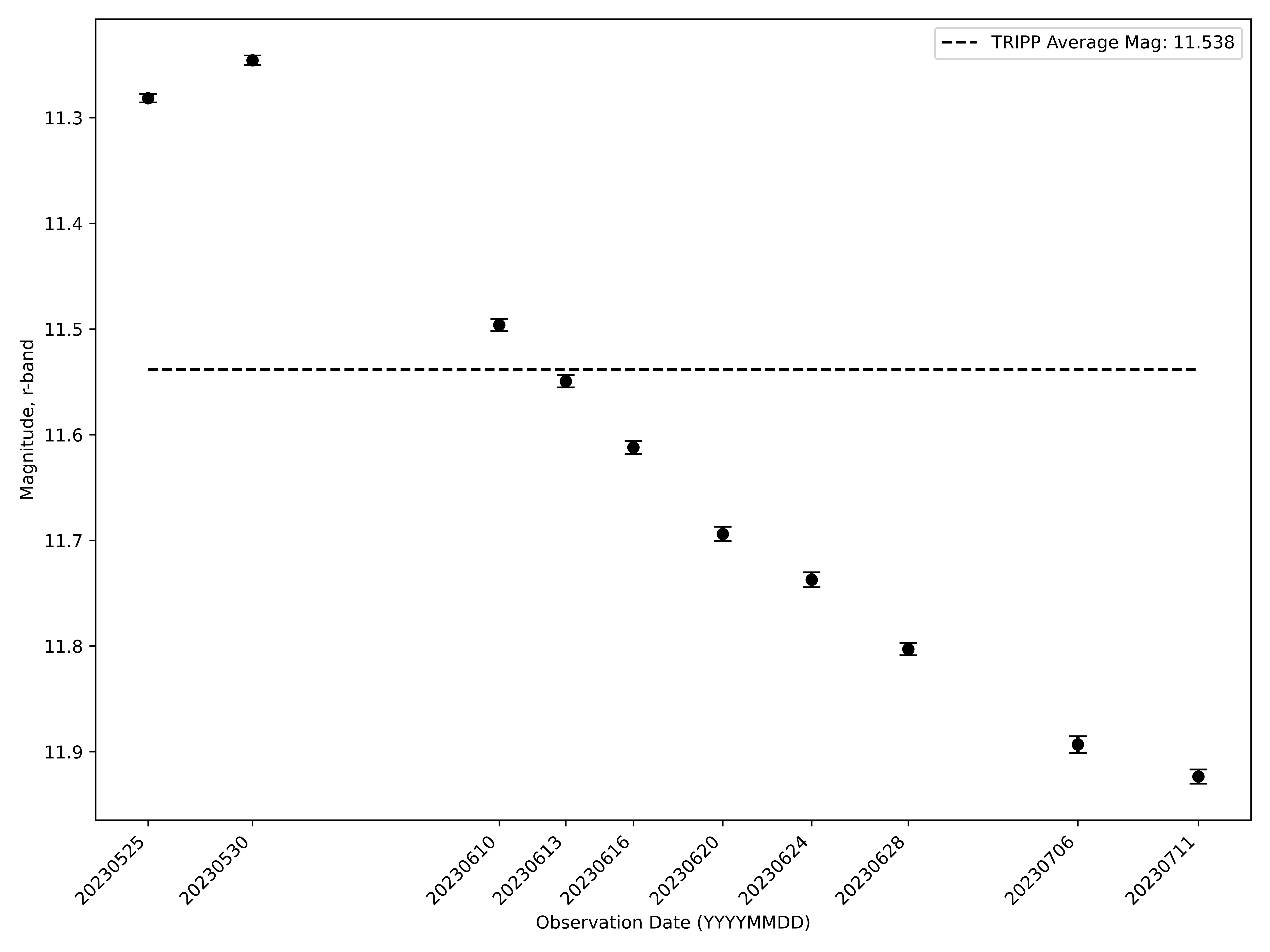}
    \caption{Light curve taken of SN2023ixf in SDSS-r' band. Each observation consists of twenty-five 15 second exposures. TRIPP's systematic error is 10 mmag.}
    \label{source34 light curve}
\end{figure}

Finally, we report a light curve of SN2023ixf taken over two months in Figure \ref{source34 light curve}. The magnitudes reported by TRIPP match those reported by \citet{Hosseinzadeh_2023}, \citet{Teja_2023}.

\section{Future of the Project}\label{FotP}
\noindent Motivated by our desire to further reduce processing time and improve accuracy, we are testing machine learning (ML) based methods of source extraction as a first step towards a primarily ML pipeline.
We are currently testing and training a neural network to obtain more accurate and faster transient detection. For this approach, we use a convolutional neural network called You Only Look Once (YOLO) Darknet, a near real-time object detection system documented in \cite{Bochkovskiy2020} and \cite{darknet13} which operates in less than 25 ms per image. 

YOLO is pretrained in the deeper layers on a large dataset of labeled everyday images compiled by Imagenet (\cite{Deng2009}).
The final layers of this second network can then be trained on a much smaller amount of specialized data to fine-tune the algorithm to the desired task of the new training.
Thus, the demand for labeled training data specific to the task of interest is reduced drastically.

We can further improve our baseline model using quick (smaller sets) retraining with images including nebulae, galaxies, and other deep-sky objects to allow our model to detect transients within noisy backgrounds. Once fully developed, evaluated, and compared to our current methods, machine learning extraction could replace the traditional Source Extractor in TRIPP and other DIA pipelines.

\section{Conclusion} \label{conclusion}
\noindent TRIPP is a real-time image analysis pipeline for the express purpose of transient detection and light curve analysis. On our hardware (see Table \ref{table:timing}), TRIPP achieves reliable \textless3 second image analysis for 6 megapixel images. We report complementary observations and analysis of SN2023ixf in agreement with previous works (RECITE). In simulated images, we report an $F_1$ statistic for transient candidates of 0.19 utilizing a threshold of the source exceeding 5$\mu_{bkg}$ in the residual image as the threshold for classification. Due to this limitation of Source Extractor to differentiate artifacts from transients, we are exploring machine learning extraction, as previously discussed, which boasts a preliminary $F_1$ statistic of 0.72.

We are presently utilizing TRIPP to detect Milky Way transients obscured by the high surface brightness of the Andromeda galaxy along with bright transients from the Andromeda galaxy. These results validate pipeline performance at a low signal-to-noise.
In addition, as TRIPP is capable of detecting optical SETI transients from civilizations with lasers that we can construct within the century (see \cite{LGTS}, \cite{Lubin2016}), Andromeda is an ideal target due to the large number of possible messenger civilizations and relative proximity.

TRIPP's general-purpose modular design is broadly employable across a variety of observation and hardware requirements from amateur astronomy to flagship observatories. 
We hope the open-source tools developed as part of this program will benefit the Time-domain Astrophysics community.

\section*{acknowledgements}
We would like to thank the many undergraduates who worked on this project before us.
In addition, TRIPP was tested using observations from the Las Cumbres Observatory Global Telescope Network, and we are grateful for the support of the network and their generous allotments of observing time. 
TRIPP utilizes a variety of open-source astronomical science data programs to create our differencing imaging pipeline within a Python framework.
These include Astroalign (\cite{Beroiz2019}), Optimal Image Subtraction (OIS; \cite{Alard1998}), Adaptive Bramich method (\cite{Miller2008}), NumPy (\cite{numpy}), and Source Extractor Python (SEP; \cite{Barbary2016}).
We also acknowledge the NVIDIA Applied Research Accelerator Program for the NVIDIA Academic Hardware Grant Award.
The processing power provided by this program in the form of the NVIDIA RTX A6000 greatly accelerated our image processing time both with the computational power and 48 GB of VRAM enabling our research to use the faster CuPy backend which would not have been possible without this support.
Funding for this project comes from the UCSB Faculty Research Assistance Program and  Undergraduate Research and Creative Activities grants, NASA grants: NIAC Phase I DEEP-IN – 2015 NNX15AL91G and NASA NIAC Phase II DEIS – 2016 NNX16AL32G, the NASA California Space Grant (NASA NNX10AT93H), as well as a generous gift from the Emmett and Gladys W. fund.

\bibliographystyle{AASJournal}
\bibliography{Bib.bib} 
\end{document}